\documentclass[twocolumn,english,showpacs,preprintnumbers,amsmath,amssymb,superscriptaddress]{revtex4}
\usepackage[T1]{fontenc}
\usepackage[latin9]{inputenc}
\usepackage{amsmath}
\usepackage{amssymb}
\usepackage{esint}

\makeatletter
\@ifundefined{textcolor}{}
{%
 \definecolor{BLACK}{gray}{0}
 \definecolor{WHITE}{gray}{1}
 \definecolor{RED}{rgb}{1,0,0}
 \definecolor{GREEN}{rgb}{0,1,0}
 \definecolor{BLUE}{rgb}{0,0,1}
 \definecolor{CYAN}{cmyk}{1,0,0,0}
 \definecolor{MAGENTA}{cmyk}{0,1,0,0}
 \definecolor{YELLOW}{cmyk}{0,0,1,0}
}

\@ifundefined{definecolor}{\usepackage{color}}{}
\usepackage{amsfonts}%%\usepackage{ascmac}
\usepackage{fancyhdr}\usepackage{fancybox}\usepackage{ulem}

\usepackage{float}%%%theoremlike environments

\usepackage{amsthm}

\def\loweq@align#1#2{\lower.6ex\vbox{\baselineskip\z@skip\lineskip\z@
    \ialign{$\m@th#1\hfil##\hfil$\crcr#2\crcr=\crcr}}}
\def\lowsim@align#1#2{\lower.6ex\vbox{\baselineskip\z@skip\lineskip\z@
    \ialign{$\m@th#1\hfil##\hfil$\crcr#2\crcr\sim\crcr}}}
\def\geqq{\mathrel{\mathpalette\loweq@align >}}
\def\leqq{\mathrel{\mathpalette\loweq@align <}}
\def\grsim{\mathrel{\mathpalette\lowsim@align >}}
\def\lesssim{\mathrel{\mathpalette\lowsim@align <}}
\def\gsim{\mathrel{\mathpalette\lowsim@align >}}
\def\lsim{\mathrel{\mathpalette\lowsim@align <}}

\newcommand{\grless}{ {\, \raise-.24em\hbox{$<$} \hspace{-0.8em} \raise.31em\hbox{$>$}\, } }
\newcommand{\lessgr}{ {\, \raise-.24em\hbox{$>$} \hspace{-0.8em} \raise.31em\hbox{$<$}\, } }
\newfont{\bg}{cmr10 scaled\magstep4}                    
\newcommand{\bigzerou}{\smash{\lower1.7ex\hbox{\bg 0}}}

\newcommand{\crl}[1]{[-\infty,\infty]}

\usepackage{babel}

\makeatother

\usepackage{babel}
\begin{document}
\pagecolor{white}

\title{Time Optimal Quantum Control of Spinor States}

\author{Peter G. Morrison}

\email{nanoscope@outlook.com}

\selectlanguage{english}%

\affiliation{Environmental Control, Sydney, Australia}

\date{July 8th, 2019}
\begin{abstract}
\noindent An analysis of the motion of a relativistic electron under
a linear constraint in four dimensions is presented. Interesting results
are given that show that the state of the electron is well defined
under the formalism of time optimal quantum state control. We establish
compact mechanisms for achieving time dependent unitary evolution,
and present new calculation methods for time-ordered exponential operators.
A powerful modification of the brachistochrone technique is presented
that allows solution of a class of problems via matrix decomposition
of the Hamiltonian. These techniques allow us to arrive at a series
of solutions for quantum systems that have readily accessible physical
realisations. We contrast the output of the theory when constrained
to a relativistic space-time to that of other physical systems of
lower dimensionality. Some comment is given regarding hypercomplex
numbers and their application in quantum mechanics.


\end{abstract}


\maketitle

\section{Introduction}

The growing need for effective methods of control in quantum systems
becomes more of a critical concern the closer we get to atomic scale
fabrication and technology. Quantum control is essential to our ability
to add up in the nanoscale domain if we are to have an effective mechanism
of computation in this realm. That we achieve the manipulation of
quantum states in a timely fashion is almost as pressing a need in
terms of a functional device.

We have set up this demonstration of time optimal quantum control
in order to show how more complicated examples in time optimisation
can be rendered into a manageable form. In this contrived environment,
we have achieved success. The approach demonstrated will show a generally
applicable approach that can be used to solve almost any dynamic system
of this type. Recognition must also be given to the development of
this new field. It is pleasing to see such a rich calculus emerge
from evidently simple constructions. Most of modern quantum mechanical
calculation starts with the basic assumptions that the system in consideration
is time independent, and also that the time dependent interaction,
if present at all, is weak in comparison to the static fields driving
the transitions between states. We shall assume neither of these properties,
and shall be examining the analytical properties of strongly driven,
periodic quantum systems where the time dependence and periodicity
is the principal part of the dynamics. These systems are of the autonomous
Floquet-type which naturally live on a finite multidimensional Hilbert
space. They have extremely interesting properties, as we shall demonstrate
throughout this paper.

The method of calculation presented will appear complementary to current
techniques; we are solving for the system Hamiltonian dependence in
time which optimises an action principle, in this case the time taken
for a state to evolve from one place to another on a complex projective
space under some simple constraints. Once we have achieved this, we
then go on to derive all other properties of the quantum system. This
is to be held in direct contrast with much of current quantum theory,
which assumes the existence of certain Hamiltonian structures. We
shall not be making any assumptions about the functional form of the
Hamiltonian as it depends on time; instead, we allow it to vary, and
find the optimal choice given the constraints at hand, the preservation
of time being one of the primary considerations of our objective.
The underlying parameter dependence emerges naturally from the consideration
of certain simple dynamical considerations which are easily evaluated
using a method of isometric transformation. That the solutions obtained
are of succinct and elegant form is obviously a consequence of the
simplicity of the chosen scenarios within which we apply this theory.
One might hope that exploration of these types of problems might allow
more of an interchange of ideas between the field of dynamical systems
and that of quantum mechanics, to look more closely at the idea of
truly dynamic quantum systems. It certainly simplifies the situation
mathematically and numerically, and aids with the physical understanding
of the systems under consideration.

In their original series of papers relating to time optimal quantum
control, Carlini et. al. addressed the control of a quantum system
in order to minimise time taken for state-to-state transfer on a complex
projective space given fixed energy and subject to linear constraints.
This was originally achieved using a brachistochrone \cite{Carlini 1},
then extended to unitary operators \cite{Carlini 2}, mixed states
\cite{Carlini 3} and Ising chains \cite{Carlini 4}. They derive
a quantum brachistochrone from an action principle which enforces
the dynamical laws of quantum evolution and the constraints which
the Hamiltonian of the physical system has to satisfy. They have further
demonstrated the general applicability of this technique, showing
that it is quite natural to use this formalism to cover a wide range
of quantum control problems, including open systems with thermodynamic
considerations \cite{Mukherjee}, transfer of state coherence \cite{Carlini 5}
and the production of time optimal gates for quantum computing and
storage procedures for special states. We note that the method contained
within this series of papers struggles at times to produce the answers
required, especially when it comes to unitary operators. The method
can be complex, and we will be seeking to modify the technique to
improve ease of application. 

The reader is directed towards worked calculations on time optimal
quantum state control for SU(2) and SU(3) \cite{Morrison 1}, where
the calculations in \cite{Carlini 1,Carlini 2} are applied and expanded
to some exotic systems of higher dimension, expanded upon in \cite{Morrison 3},
where eigenvalue methods were used to examine unitary transformations
on a number of problems. In particular, this paper develops some observations
considered in \cite{Morrison 2} relating to the optimal control of
a time dependent biqubit system.

For computer applications, the reader is referred to \cite{Wang}
where some numerical methods are examined that can be used to solve
the quantum brachistochrone equation. The authors observe performance
problems and scaling issues even for moderately large dimensions.
Computationally, this is related to calculating the brachistochrone
over all possible subcombinations of Hamiltonian and constraint to
find globally time optimal solutions.

As for other examples of physical systems where time optimal control
techniques may be applied, we direct towards \cite{Jirari} for an
outline regarding a DC squid problem given applied time-dependent
bias current and magnetic flux for a driven multi-level quantum system.
The use of spectral filter theory in this domain allows both a place
to find possible continuous state formalisms for time optimal control
theory as well as an experimental testing ground. 

Optimisation theory has been applied for many years within the field
of photochemical control. \cite{Gross} contains an outline of photochemical
control using radiative coupling; this is used to control the output
of the chemical reaction, using mathematical techniques that are structurally
similar to our time optimal control problem on SU(2). The functionals
used as objectives for the optimisation procedure differ from the
methods of time optimal control; in this field one is primarily focused
on variation of the transmitted probability of an electronic state
in order to maximise yields of product. Other references related to
photochemistry may be found in \cite{Somloi}for the photochemical
control of $I_{2}$ dissociation, also using optimal control to manipulate
transition probabilities, and providing a test ground for SU(2) problems.
Unitary operators that target overlaps were also considered in \cite{Palao}
where diatomic surfaces \& chemical control were used to develop a
quantum Fourier transform.

We note the large experimental and theoretical literature devoted
to gradient descent methods. These developments have led the way through
exploring the role time plays in spin-NMR optimal control, e.g. in
\cite{Khaneja 1}, where techniques of sub-Riemannian geometry as
applied to the unitary evolution of the quantum state were given,
also \cite{Khaneja 2} where SO(3) algebra was used to solve a spin
system, and \cite{Khaneja 3} for gradient descent methods. Other
research relating to magic-angle transformations may be found in \cite{Vosegaard,Schulte-Herbruggen}.
Papers relating to sub-Riemannian geometry and other aspects of quantum
state control may be found in \cite{Brockett 1,Brockett 2}, where
a critical distinction is made between open and closed loop formalisms.
Observations are made relating to the relationship between the Riemannian
structure on a manifold, its natural measure, and how to obtain an
invariant operator which may prove useful for the continuous case.
For experimental description of gradient descent in spin magentic
resonance systems, consult \cite{Chen}. In \cite{Kehlet}, coherence
transfer efficiences were increased via optimal control sequences
using inhomogenous rf fields. Further, \cite{Li} examined a hybridised
version of quantum computing where the classical computer serves a
role in a gradient-based optimal control system. 

The applications for this type of optimisation problem are not solely
within quantum mechanics; other authors within operations research
and dynamical programming have examined related topics. In particular,
\cite{Fattorini} c. 1964 looks at some properties of time dependent
dynamic systems that have similarities to the systems we have considered,
while \cite{Asarin} tackles the time optimal control problem from
another direction using automata and computer simulation. \cite{Bobrow}
outlines a technique to minimise the time taken for a discrete system
for a robotic manipulator, whereas \cite{Lee} also examined autonomous
control systems of a discrete nature. Other earlier moves to study
the problem of discrete time controllers for dynamic systems under
constraints may be found in \cite{Mayne}; finally, \cite{Friesz}
looked at the time optimisation problem for a system of networked
traffic to try and find optimised flows.

The need for concrete examples and concise notation is recognised,
especially given how complicated the following calculation quickly
becomes. An effort has been made to reduce the load of dealing with
such intricate matrix calculus and non-commutative objects. For that
reason, every opportunity has been taken to simplify notation and
give explicit descriptions of the objects used throughout the calculation
where necessary. As we are using objects in general that can't be
assumed to commute, often the order of multiplication will be defined
in full. Extra levels of difficulty are added through the use of complex
matrix identities that form part of underlying blocks of matrices
that are of higher dimensions. This will be shown to be a consequence
of certain transformation properties of unitary operators that describe
the time evolution of the quantum system.

Our mathematical apparatus shall exploit matrices of a particular
type well known to physics, originally explored in \cite{Dirac}.
In \cite{Redkov}, the authors develop spinor transformations and
unitary operators on SU(4) which have similar properties to what we
derive in this paper. The major results we shall focus on include
an Euler angle decomposition, and the extraction of the SU(3) subgroups
of SU(4). In particular, the unitary operators, composition laws of
the group and formulae for the inversion of matrices they develop
are mathematically similar to our findings. We shall have to resort
to clever transformations, and the understanding of chirality and
twisted states in the Dirac spinor space is useful. The reader is
directed towards \cite{Shi}, where chiral states can be used and
their relationship to spin \& helicity, with applications to muon
decay. For other types of systems that explore similar mathematical
techniques, refer to \cite{Mandal}, who looks at the spin Hall effect
and Rashba Hamiltonians, where time optimal state control might be
of interest. We shall have particular use for the methods of Floquet
\cite{Floquet}. Notes with particular relevance to two level systems
and the interaction of polarised states via Floquet theory may be
found in \cite{Santoro}.

For application to curvilinear coordinates to the Dirac equation,
consult \cite{Leclerc}, where some equations which may be related
to the quantum brachistochrone equation considered in \cite{Carlini 1}
are derived. Other systems related to the calculation in this paper
can be found in \cite{Rodionov}, where some non-Hermitian modifications
of the Dirac-Pauli equations are considered. 

As we shall have some use in passing for SU(3), a descriptor of SU(3)
symmetry states and their implementation in beryllium isotopes is
contained in \cite{Almeida} where an SU(3) form of quadrupolar-quadrupolar
interaction is used to model a set of excitations. This may provide
us with a mechanism to move to continuous systems via orthogonal polynomials.
\cite{Sklarz} looks at some more complex SU(3) problems using a canonical
decomposition,they calculated in order to achieve state control using
the Hamilton-Jacobi-Bellman equation. Finally, in \cite{Lapert} ,
some problems similar in dimension to the SU(3) case are addressed
in full.

The outline of the paper will be as follows; in section II we outline
the matrix calculus we will be using in the problem, III will address
the quantum control of the system and some simple dynamical systems
calculations, IV will apply a transformation of eigenstates, and V
will discuss the quantum brachistochrone equation. We will then use
the results obtained to evaluate the time evolution operator of the
system in VI, and discuss results for some related problems on other
dimensions in VII. Finally, we shall discuss some ways in which the
quantum brachistochrone method could be extended in VIII, and examine
some interesting statements from the field of quantum electrodynamics
which this calculation has addressed.

\section{Physical Considerations}

Consider the standard Dirac Hamiltonian matrix for a spinning electron,
using natural units:

\begin{flushleft}
\begin{equation}
\tilde{H}(t)=\mathbf{\hat{\mathbf{\alpha}}}\centerdot\mathbf{p}+\hat{\beta}m
\end{equation}
We use bold lowercase letters to represent a vector such as $\mathbf{p}=(p_{x},p_{y},p_{z})$,
with matrix operators to be represented by the caret as with $\hat{\beta}$,
and a vector of matrices to be given by a bolded caret as with$\hat{\mathbf{\alpha}}=(\hat{\alpha}_{x},\hat{\alpha}_{y},\hat{\alpha}_{z})$.
We choose the particular basis given by Kronecker products of the
Pauli matrices and the identity matrix $\hat{\beta}=\hat{\sigma}_{z}\otimes1$,
$\hat{\alpha}_{j}=\hat{\sigma}_{x}\otimes\hat{\sigma}_{j}$ so that
the Hamiltonian is Hermitian. One can immediately write down the standard
rules of the algebra, in this case we work with the Hermitian anti-commuting
quartic roots of unity, $\hat{\alpha}_{j}^{\dagger}=\hat{\alpha}_{j}$,
$\hat{\beta}^{\dagger}=\hat{\beta}$, \& similarly:
\par\end{flushleft}

\begin{equation}
\hat{\beta}^{2}=\hat{\alpha}_{j}^{2}=\mathbf{1}
\end{equation}
\begin{equation}
\left\{ \hat{\alpha}_{j},\hat{\alpha}_{k}\right\} =2\delta^{jk}
\end{equation}
\begin{equation}
\left\{ \hat{\beta},\hat{\alpha}_{k}\right\} =0
\end{equation}
This ensures that the Hamiltonian is completely Hermitian and has
the simple matrix form displayed below:

\begin{equation}
\tilde{H}=\left[\begin{array}{cc}
m\mathbf{1} & -i(\mathbf{p}\centerdot\hat{\mathbf{\sigma}})\\
i(\mathbf{p}\centerdot\hat{\mathbf{\sigma}}) & -m\mathbf{1}
\end{array}\right]
\end{equation}
We have the standard time evolution equation for the time dependent
state $i\dfrac{d}{dt}\left|\Psi(t)\right\rangle =\tilde{H}(t)\left|\Psi(t)\right\rangle $.
We shall not use partial differentiation in this paper, so all derivatives
will be with respect to the underlying time parameter. The tilde over
an operator indicates that it is trace-free and Hermitian, which is
true by observation. We could write this for the constituent operators
but this is not necessary; we shall conform to the notation as originally
intended.

Using an action principle of
\begin{equation}
S=\int1dt+\int\lambda_{1}\textrm{Tr}\left(\dfrac{\tilde{H}^{2}}{2}-E^{2}\right)dt+\int\lambda_{2}\textrm{Tr}\left(\tilde{H}\tilde{F}\right)dt
\end{equation}
we may minimise the time taken between states on the complex projective
space via the Fubini-Study metric:
\begin{equation}
1dt=\dfrac{1}{\Delta E}\sqrt{(\left\langle d\Psi\right|\left(\mathbf{1}-\left|\Psi\right\rangle \left\langle \Psi\right|\right)\left|d\Psi\right\rangle )}
\end{equation}
originally derived in \cite{Carlini 1}, expanded upon in \cite{Carlini 2,Morrison 1},
where we implicitly assume the state evolves according to the Schrödinger
equation as $id\left|\Psi\right\rangle =\tilde{H}(t)\left|\Psi\right\rangle dt$,
the energy dispersion being defined by $\Delta E(t)=\left\langle \Psi\right|\tilde{H}^{2}(t)\left|\Psi\right\rangle -\left(\left\langle \Psi\right|\tilde{H}(t)\left|\Psi\right\rangle \right)^{2}$,
and the action principle given by the standard formula $\delta S=0$.
The Lagrange multipliers, assumed to be equal zero at all times, represent
constraints on the physical system such that it is free to explore
some dimensions of the system, while being constrained in others,
and held to a condition of finite energy. For further discussion and
the complete original argument in detail, the reader should consult
\cite{Carlini 1}. The implicit assumptions for the dynamics of the
system may seen as the addition of a further Lagrangian to the action
principle in order to constrain the dynamics as in \cite{Carlini 1},
however, as we will not have need for the extension to mixed states,
it is sufficient for our purpose to simply note that the result is
the Heisenberg equation for certain operators and some complicated
relations for boundary conditions. In some ways, this is the perfect
application for this theory, as relativity implies constraints on
the relationship between energy, momentum and rest mass as per the
relations of Einstein, which imply the existence of certain types
of metric spaces. We shall see through the series of examples that
follows that we are able to satisfy all the above demands and more.

\section{Quantum Control}

Let us now consider the quantum control equations. As $\tilde{H}$
is Hermitian, and trace-free, we can form another matrix independent
of this such that $\textrm{Tr}\left(\tilde{H}\tilde{F}\right)=0$
using a decomposition over the Hermitian, trace-free generators of
the group. We shall call the matrix operator $\tilde{F}(t)$ as 'constraint';
it limits the degrees of freedom that the system has access to over
the whole operator space. This matrix will also be Hermitian and trace-free
by construction. Using the Heisenberg equation of motion $i\dfrac{d\hat{A}}{dt}=[\tilde{H}(t),\hat{A}]$,
and forming the operator $\hat{A}=\tilde{H}+\tilde{F}$, we can derive
the following equation

\begin{equation}
i\dfrac{d}{dt}\left(\tilde{H}+\tilde{F}\right)=[\tilde{H},\tilde{H}+\tilde{F}]=[\tilde{H},\tilde{F}]=\tilde{H}\tilde{F}-\tilde{F}\tilde{H}
\end{equation}
where we have used the fact that the Hamiltonian matrix commutes with
itself at a fixed time. In a sense this derivation is naïve, and only
proves consistency, without sufficiency; it assumes the Heisenberg
equation of motion, to be contrasted with using the analysis of the
brachistochrone in \cite{Carlini 1}.

The final piece of data required for the analysis of this problem
is to calculate the isotropic constraint itself. Calculating $\textrm{Tr}\left(\tilde{H}^{2}/2\right)=m^{2}+\left|\mathbf{p}\right|^{2}=\textrm{const.}<\infty$,
we find that for the quantum control theory to be valid, we must have
the state constrained to the surface of a sphere of finite size, in
this case in four dimensions. We will write this explicitly in the
form $E^{2}=m^{2}+\left|\mathbf{p}\right|^{2}$ for some constant
$E$; note also that we have the result $\tilde{H}^{2}=E^{2}\mathbf{1}$,
analogous to the Klein-Gordon equation, proven using the matrix formula
in eq. (5). 

The calculation of the equations in (8) can be simplified using the
following technique. First, break down the Hamiltonian matrix and
constraint as sums over the generators of the group via $\tilde{H}=\sum^{'}\lambda_{i}\hat{g}_{i}$,
$\tilde{F}=\sum^{''}\varOmega_{j}\hat{g}_{j}$, where the sum in the
Hamiltonian is over the set $S$ and that of the constraint is over
$S^{C}$, i.e. we have split the whole space into two sections, termed
Hamiltonian and constraint. We can then write the quantum brachistochrone
equation (6) in the form:

\begin{equation}
i\left(\sum_{S}\dfrac{d\lambda_{i}}{dt}\hat{g}_{i}+\sum_{S^{C}}\dfrac{d\Omega_{j}}{dt}\hat{g}_{j}\right)=\sum_{S}\sum_{S^{C}}\lambda_{i}\Omega_{j}\left[\hat{g}_{i},\hat{g}_{j}\right]
\end{equation}
Using the orthogonality property of the generators of the space, we
have $\textrm{Tr}\left[\hat{g}_{k}\hat{g}_{l}\right]=\delta_{kl}$,
so by multiplying by an element of the group $\hat{g}_{k}$ and computing
the trace of eq.(9) one can read off the independent components of
the dynamical equations. At this point, we display the constraint
matrix explicitly as:

\begin{equation}
\tilde{F}=\left[\begin{array}{cccc}
\Omega_{+} & \xi_{01} & \xi_{02} & \xi_{03}\\
\xi_{01}^{*} & -\Omega_{+} & \xi_{12} & \xi_{13}\\
\xi_{02}^{*} & \xi_{12}^{*} & \Omega_{-} & \xi_{23}\\
\xi_{03}^{*} & \xi_{13}^{*} & \xi_{23}^{*} & -\Omega_{-}
\end{array}\right]=\sum_{i,j\in S^{C}}\Omega_{ij}(t)[\hat{\sigma}_{i}\otimes\hat{\sigma}_{i}]
\end{equation}
where we have defined $\Omega_{\pm}=\Omega_{33}+\Omega_{03}$, also
$\hat{\sigma}_{0}=\mathbf{1}_{2\times2}$ and the complex coefficients
as below:
\begin{equation}
\begin{array}{c}
\xi_{01}=\Omega_{31}-i\Omega_{32}+\Omega_{01}-i\Omega_{02}\\
\xi_{23}=-\Omega_{31}+i\Omega_{32}+\Omega_{01}-i\Omega_{02}\\
\xi_{13}=\Omega_{10}-i\Omega_{20}+i\Omega_{23}\\
\xi_{02}=\Omega_{10}-i\Omega_{20}-i\Omega_{23}\\
\xi_{03}=-\Omega_{22}-i\Omega_{21}\\
\xi_{12}=\Omega_{22}-i\Omega_{21}
\end{array}
\end{equation}
This choice of constraint matrix guarantees the relationship $\textrm{Tr}\left(\tilde{H}\tilde{F}\right)=0$.
Note that we might equally have chosen the representation of the space
in terms of ascending products of Dirac gamma matrices. An example
calculation is shown below:

\begin{equation}
i\dfrac{d}{dt}\left(\textrm{Tr}\left[\hat{\beta}\left(\tilde{H}+\tilde{F}\right)\right]\right)=\textrm{Tr}\left[\hat{\beta}\left[\tilde{H},\tilde{F}\right]\right]
\end{equation}
\begin{equation}
\dfrac{dm}{dt}=2\left(\Omega_{21}p_{x}+\Omega_{22}p_{y}+\Omega_{23}p_{z}\right)
\end{equation}
Computing this operation over all elements of the group $\hat{\sigma}_{i}\otimes\hat{\sigma}_{j}$,
we obtain the following set of first-order differential equations:

\begin{equation}
\dfrac{d\Omega_{0j}}{dt}=\dfrac{d\Omega_{2j}}{dt}=0,j=1,2,3
\end{equation}
\begin{equation}
\dfrac{d}{dt}\left[\begin{array}{c}
\Omega_{10}\\
\Omega_{31}\\
\Omega_{32}\\
\Omega_{33}
\end{array}\right]=2\left[\begin{array}{cccc}
-1 & 0 & 0 & 0\\
0 & 1 & 0 & 0\\
0 & 0 & 1 & 0\\
0 & 0 & 0 & 1
\end{array}\right]\left[\begin{array}{c}
m\\
p_{x}\\
p_{y}\\
p_{z}
\end{array}\right]
\end{equation}
\begin{equation}
\dfrac{d}{dt}\left[\begin{array}{c}
m\\
p_{x}\\
p_{y}\\
p_{z}
\end{array}\right]=2\left[\begin{array}{cccc}
0 & \Omega_{21} & \Omega_{22} & \Omega_{23}\\
-\Omega_{21} & 0 & \Omega_{03} & -\Omega_{02}\\
-\Omega_{22} & -\Omega_{03} & 0 & \Omega_{01}\\
-\Omega_{23} & \Omega_{02} & -\Omega_{01} & 0
\end{array}\right]\left[\begin{array}{c}
m\\
p_{x}\\
p_{y}\\
p_{z}
\end{array}\right]
\end{equation}
\begin{equation}
\dfrac{d\Omega_{20}}{dt}=2(m\Omega_{10}-p_{x}\Omega_{31}-p_{y}\Omega_{32}-p_{z}\Omega_{33})
\end{equation}
This can be succinctly summarised in vector-matrix notation as $\mathbf{\dot{\Omega}_{0}=\dot{\Omega}_{2}=0}$,
$\dot{\Xi}_{\mu}=-S_{\mu\nu}p_{\nu}$ , $\dot{p}_{\mu}=\varTheta_{\mu\nu}p_{\nu}$
and $\dot{s}=p_{\mu}S_{\mu\nu}\Xi_{\nu}$. Additionally, we have that
$\varTheta_{\mu\nu}=-\varTheta_{\nu\mu}$ by observation. This system,
while complex, is solvable. However, we have one principal difficulty
in that the momentum equation contains a matrix that is not easily
exponentiated, given that we are not provided with the initial values
$\Omega_{2j}$ and $\Omega_{0j}$. At this point, it appears that
we are at a stop. It is important that the number of additional assumptions
is reduced, if at all possible, and the boundary conditions of the
theory should emerge naturally rather than as an extra artificial
constraint. The next section shall demonstrate that considerable simplification
of this complicated set of coupled differential equations may be achieved
by looking at the dynamics of the system in a transformed reference
frame. This will render the need for extra boundary conditions placed
upon the initial and terminal values on the quantum state to be superfluous
to the correct implementation of a time optimal unitary evolution.

\section{Matrix of Eigenstates}

Given the matrix $\tilde{H}$ defined in eq. (5) we can write a simple
eigenvalue equation $\tilde{H}\left|\mathbf{n}(t)\right\rangle =E_{n}\left|\mathbf{n}(t)\right\rangle $for
the generally time-dependent eigenstates. As this section deals almost
solely with time-dependent frames of reference, we shall drop the
explicit time dependence on the states $\left|\mathbf{n}(t)\right\rangle $and
just write$\left|\mathbf{n}\right\rangle $. Consider now the matrix
that is given by the set of eigenstates:

\begin{equation}
\hat{W}=\left[\begin{array}{cccc}
\vdots & \vdots & \vdots & \vdots\\
\left|\mathbf{u}_{1}\right\rangle  & \left|\mathbf{u}_{2}\right\rangle  & \left|\mathbf{v}_{1}\right\rangle  & \left|\mathbf{v}_{2}\right\rangle \\
\vdots & \vdots & \vdots & \vdots
\end{array}\right]
\end{equation}
and an inverse matrix of eigenstates as below:
\begin{equation}
\hat{W}^{-1}=\left[\begin{array}{ccc}
\cdots & \left\langle \bar{\mathbf{u}}_{1}\right| & \cdots\\
\cdots & \left\langle \bar{\mathbf{u}}_{1}\right| & \cdots\\
\cdots & \left\langle \bar{\mathbf{v}}_{1}\right| & \cdots\\
\cdots & \left\langle \bar{\mathbf{v}}_{2}\right| & \cdots
\end{array}\right]
\end{equation}
We must now address the question of how to produce $\left\langle \bar{\mathbf{u}}_{1}\right|$
from $\left|\mathbf{u}_{1}\right\rangle $. Firstly, we can derive
the eigenvalue equation from $\det\left(\tilde{H}(t)-\lambda\mathbf{1}\right)=0$.
We obtain the following polynomial
\begin{equation}
\left(\lambda^{2}-E^{2}\right)^{2}=0
\end{equation}
with $E=\pm\sqrt{m^{2}+\left|\mathbf{p}\right|^{2}}$, which is independent
of time. We must therefore have two eigenvalues, each with multiplicity
two. We can write the eigenvalue equation for the left and right forms
of the eigenstate matrix as $\tilde{H}\hat{W}=E\hat{L}\hat{W}$, conversely
$\hat{W}^{-1}\tilde{H}=E\hat{W}^{-1}\hat{L}$, where we have the matrix
$\hat{L}$ given explicitly as:

\begin{equation}
\hat{L}=\left[\begin{array}{cccc}
1 & 0 & 0 & 0\\
0 & 1 & 0 & 0\\
0 & 0 & -1 & 0\\
0 & 0 & 0 & -1
\end{array}\right]=\hat{\sigma}_{z}\otimes\mathbf{1}
\end{equation}
Note that $\tilde{H}^{-1}=\tilde{H}/E^{2}$ by the matrix identity
$\tilde{H}^{2}=E^{2}\mathbf{1}$ and that all matrices in the above
equations are invertible. We can construct the eigenmatrices: 
\begin{equation}
\hat{W}=\left[\begin{array}{cccc}
\dfrac{p_{x}-ip_{y}}{E-m} & \dfrac{p_{z}}{E-m} & -\dfrac{(p_{x}-ip_{y})}{E+m} & \dfrac{-p_{z}}{E+m}\\
\dfrac{-p_{z}}{E-m} & \dfrac{(p_{x}+ip_{y})}{E-m} & \dfrac{p_{z}}{E+m} & -\dfrac{(p_{x}+ip_{y})}{E+m}\\
0 & 1 & 0 & 1\\
1 & 0 & 1 & 0
\end{array}\right]
\end{equation}
\begin{equation}
\hat{W}^{-1}=\dfrac{1}{2E}\left[\begin{array}{cccc}
p_{x}+ip_{y} & -p_{z} & 0 & E-m\\
p_{z} & p_{x}-ip_{y} & E-m & 0\\
-(p_{x}+ip_{y}) & p_{z} & 0 & E+m\\
-p_{z} & -p_{x}+ip_{y} & E+m & 0
\end{array}\right]
\end{equation}
where we also have the implicit equation $E^{2}=m^{2}+\left|\mathbf{p}\right|^{2}$,
which when used together demonstrates that $\hat{W}\hat{W}^{-1}=\hat{W}^{-1}\hat{W}=\mathbf{1}$
as required. It is immediately apparent that we do not have $\hat{W}^{-1}=\hat{W}^{\dagger}$,
so for this system we must work very carefully as it is not unitary
even if it is invertible. These formulae are all valid for any energy
and momentum values, as long as we maintain quantum state separation,
via $E\neq0$. In particular, they are true for the situation in which
the momentum values $p_{j}$ and mass might explicitly depend on the
time parameter.

\section{Quantum Brachistochrone Equation}

Consider an isometric transformation of the Hamiltonian via $\hat{D}_{0}=\hat{W}^{-1}\tilde{H}\hat{W}$.
For our particular example, using $\hat{W}(t)$, $\hat{W}^{-1}(t)$
as above, where we are now labelling the time dependence explicitly,
we find the simple expression:

\begin{equation}
\tilde{H}(t)=\hat{W}(t)\hat{D}_{0}\hat{W}^{-1}(t)
\end{equation}
where $\hat{D}_{0}=E\hat{L}$ as in the previous section. Let us calculate
the time-rate of change for the eigenmatrices. We have, by construction,
that $\hat{\dot{W}}=-i\tilde{H}(t)\hat{W}(t)=-i\hat{D}_{0}\hat{W}$.
Taking the derivative of the identity matrix, we find 
\begin{equation}
\dfrac{d}{dt}(\mathbf{1})=\dfrac{d}{dt}(\hat{W}^{-1}\hat{W})=\dfrac{d\hat{W}^{-1}}{dt}\hat{W}+\hat{W}^{-1}\dfrac{d\hat{W}}{dt}=0
\end{equation}
which gives us the identity $\hat{\dot{W}}^{-1}\hat{W}=-\hat{W}^{-1}\hat{\dot{W}}$,
the overdot indicating differentiation with respect to the time parameter.
The equation for the inverse eigenmatrix evolution is then $\hat{\dot{W}}^{-1}=i\hat{W}^{-1}\tilde{H}=i\hat{W}^{-1}\hat{D}_{0}$.
We can then explicitly differentiate eq. (24) to obtain 
\begin{equation}
i\dfrac{d\tilde{H}}{dt}=[\tilde{H},\hat{D}_{0}]
\end{equation}
which aids with the solution of the quantum brachistochrone equation.
Let us now directly solve the remaining expressions. We may write
the Hamiltonian in the following form:
\begin{equation}
\tilde{H}(t)=\hat{W}(t)\hat{D}_{0}\hat{W}^{-1}(t)
\end{equation}

\begin{equation}
\tilde{H}(t)=e^{i\hat{D}_{0}t}\hat{W}(0)\hat{D}_{0}\hat{W}^{-1}(0)e^{-i\hat{D}_{0}t}
\end{equation}
Computing directly, we obtain the time-optimal Hamiltonian operator:
\begin{equation}
\tilde{H}(t)=\left[\begin{array}{cc}
m\mathbf{1} & e^{-i(2Et+\theta)}\mathbf{p}(0)\cdot\mathbf{\sigma}\\
e^{i(2Et+\theta)}\mathbf{p}(0)\cdot\mathbf{\sigma} & -m\mathbf{1}
\end{array}\right]
\end{equation}
from which we conclude that $\dot{m}=0$. For the momentum component,
we may immediately write down the solution. Note that the arbitrary
phase can be taken as $\theta=-\pi/2$, in which case we recover the
optimal Hamiltonian as 
\begin{equation}
\tilde{H}(t)=\left[\begin{array}{cc}
m\mathbf{1} & -ie^{-2iEt}\mathbf{p}(0)\cdot\mathbf{\sigma}\\
ie^{2iEt}\mathbf{p}(0)\cdot\mathbf{\sigma} & -m\mathbf{1}
\end{array}\right]
\end{equation}

\begin{equation}
\tilde{H}=\left[\begin{array}{cc}
m\mathbf{1} & -i(\mathbf{p}\centerdot\hat{\mathbf{\sigma}})\\
i(\mathbf{p}\centerdot\hat{\mathbf{\sigma}}) & -m\mathbf{1}
\end{array}\right]
\end{equation}
We find automatically that $\tilde{H}^{2}=E^{2}\mathbf{1}$. Differentiating
directly, it is simple to show that 
\begin{equation}
i\dfrac{d\tilde{H}}{dt}=2E\left[\begin{array}{cc}
\mathbf{0} & e^{-2iEt}\mathbf{p}(0)\cdot\mathbf{\sigma}\\
-e^{2iEt}\mathbf{p^{\star}}(0)\cdot\mathbf{\sigma} & \mathbf{0}
\end{array}\right]
\end{equation}
Evaluating the commutator with the diagonal matrix of eigenvalues:
\begin{equation}
[\tilde{H},\hat{D}_{0}]=2E\left[\begin{array}{cc}
\mathbf{0} & -i(\mathbf{p}\centerdot\hat{\mathbf{\sigma}})\\
i(\mathbf{p^{\star}}\centerdot\hat{\mathbf{\sigma}}) & \mathbf{0}
\end{array}\right]=i\dfrac{d\tilde{H}}{dt}
\end{equation}
We have used a star to denote the complex conjugate of the momentum
vector. Although it is a vector that can be seen as a real, three-dimensional
vector, in a sense it is more correct to model it as a vector with
real $p_{z}$ and $p_{x}+ip_{y}$ rotating in the complex plane.

\section{Time Evolution Operator}

Let us now return to calculation of the time evolution operator. We
can write the time dependent Hamiltonian as a matrix decomposition:
\begin{equation}
\tilde{H}(t)=\hat{W}(t)\hat{D}_{0}\hat{W}^{-1}(t)
\end{equation}
\begin{equation}
=\left[\hat{W}(t)\hat{W}^{-1}(0)\right]\left[\hat{W}(0)\hat{D}_{0}\hat{W}^{-1}(0)\right]\left[\hat{W}(0)\hat{W}^{-1}(t)\right]
\end{equation}
\begin{equation}
=\hat{U}(t,0)\tilde{H}(0)\hat{U}^{-1}(t,0)
\end{equation}
where we define $\hat{U}(t,s)=\hat{W}(t)\hat{W}^{-1}(s)$. We must
now establish the unitary nature of the operator $\hat{U}(t,s)$ to
finish the proof. We can explicitly construct our eigenstate matrices
in the following block-diagonal format: 
\begin{equation}
\hat{W}(t)=\left[\begin{array}{ccc}
\dfrac{\mathbf{\mathbf{\epsilon}}\cdot\mathbf{p}_{0}e^{-2iEt}}{E_{-}} & \vdots & \dfrac{-\mathbf{\mathbf{\epsilon}}\cdot\mathbf{p}_{0}e^{-2iEt}}{E_{+}}\\
\cdots & \cdot & \cdots\\
\hat{\sigma}_{x} & \vdots & \hat{\sigma}_{x}
\end{array}\right]
\end{equation}
\begin{equation}
\hat{W}^{-1}(t)=\dfrac{1}{2E}\left[\begin{array}{ccc}
\mathbf{\mathbf{\epsilon^{\dagger}}}\cdot\mathbf{p}_{0}e^{2iEt} & \vdots & E_{-}\hat{\sigma}_{x}\\
\cdots & \cdot & \cdots\\
-\mathbf{\mathbf{\epsilon}^{\dagger}}\cdot\mathbf{p}_{0}e^{2iEt} & \vdots & E_{+}\hat{\sigma}_{x}
\end{array}\right]
\end{equation}
where we define the vector of matrices $\mathbf{\mathbf{\epsilon}}=(\mathbf{1},-i\hat{\sigma}_{z},+i\sigma_{y})$
for brevity. We have used the time evolution of the eigenmatrices
as $\hat{W}(t)=e^{-it\hat{D}_{0}}\hat{W}(0)$, also $\hat{W}^{-1}(t)=\hat{W}^{-1}(0)e^{it\hat{D}_{0}}$.
A simple calculation shows the matrix identity
\begin{equation}
\left(\mathbf{\epsilon}\cdot\mathbf{p}\right)\left(\mathbf{\epsilon^{\dagger}}\cdot\mathbf{p}\right)=\left(\mathbf{\epsilon^{\dagger}}\cdot\mathbf{p}\right)\left(\mathbf{\epsilon}\cdot\mathbf{p}\right)=\left|\mathbf{p}\right|^{2}\mathbf{1}_{2\times2}
\end{equation}
We can now calculate eigenmatrix properties:

\begin{equation}
\hat{W}^{-1}(t)\hat{W}(t)=\mathbf{1}
\end{equation}
where we have used $E_{+}E_{-}=\left|\mathbf{p}\right|^{2}$. Conversely,
for the opposite side of the identity, we find $\hat{W}(t)\hat{W}^{-1}(t)=\mathbf{1}$.
We are nearly complete, as this establishes the identity $\hat{U}(t,t)=\mathbf{1}$.
We have demonstrated the explicit time dependent form of the matrix
of eigenstates. We may now directly evaluate the time evolution operator
via 
\begin{equation}
\hat{U}(t,s)=\hat{W}(t)\hat{W}^{-1}(s)
\end{equation}
Direct matrix multiplication gives the final result: 
\begin{equation}
\hat{U}(t,s)=\left[\begin{array}{ccc}
e^{iE(t-s)}\mathbf{1} & \vdots & \mathbf{0}\\
\cdots & \cdot & \cdots\\
\mathbf{0} & \vdots & e^{-iE(t-s)}\mathbf{1}
\end{array}\right]
\end{equation}
where we have used $(\mathbf{\mathbf{\epsilon}}\cdot\mathbf{p})(\mathbf{\mathbf{\epsilon^{\dagger}}}\cdot\mathbf{p})=\left|\mathbf{p}\right|^{2}\mathbf{1}$,
as well as the conservation law $E^{2}-m^{2}=E_{+}E_{-}=\left|\mathbf{p}_{0}\right|^{2}$.We
manifestly have the necessary unitary properties, such as $\hat{U}^{-1}(t,s)=\hat{U}^{\dagger}(t,s)$,
also $\hat{U}^{*}(-t,-s)=\hat{U}(t,s)$. This is the major result
of this paper, as it is a new solveable system in the field of quantum
dynamics. We have succeeded in our seemingly insurmountable task of
disentangling the variables of a relativistic electron. We now, for
completeness, evaluate the dynamical system exhibited in the preliminaries.
We can firstly obtain the constraint as a function of time via the
unitary operator: 
\begin{equation}
\hat{U}(t,0)\tilde{F}(0)\hat{U}^{\dagger}(t,0)=\tilde{F}(t)
\end{equation}
\begin{equation}
\tilde{F}(t)=\left[\begin{array}{ccc}
\mathbf{\sigma}\cdot\mathbf{n}_{+} & \vdots & -e^{2iEt}\left(a\mathbf{1}-i\mathbf{\sigma}\cdot\mathbf{b}\right)\\
\cdots & \cdot & \cdots\\
e^{-2iEt}\left(a^{\star}\mathbf{1}+i\mathbf{\sigma}\cdot\mathbf{b}\right) & \vdots & \mathbf{\sigma}\cdot\mathbf{n}_{-}
\end{array}\right]
\end{equation}
where we define the split-complex variables as $a=\Omega_{10}+i\Omega_{20}$,
$\mathbf{n}_{\pm}=[\Omega_{0j}\pm\Omega_{3j}]$, $\mathbf{b}=[\Omega_{2j}]$.
Rewriting the original system of equations in vector-matrix form,
we obtain the following:
\begin{equation}
\dot{\mathbf{p}}=-m\left(\mathbf{n}_{+}+\mathbf{n}_{-}\right)-\left(\mathbf{n}_{+}+\mathbf{n}_{-}\right)\mathbf{\times}\mathbf{p}
\end{equation}

\begin{equation}
\dot{\xi}_{c}=m\xi_{r}+\mathbf{p}\cdot(\mathbf{n}_{+}-\mathbf{n}_{-})
\end{equation}

\begin{equation}
\dot{\mathbf{n}}_{+}+\dot{\mathbf{n}}_{-}=4\mathbf{p}
\end{equation}
as well as $\dot{\mathbf{n}}_{+}=\dot{\mathbf{n}}_{-}$, $\dot{\xi}_{r}=-m$,
$\dot{m}=\mathbf{b}\cdot\mathbf{p}$, $\dot{\mathbf{b}}=0$.

\section{Other Dimensions}

Given the success of this technique, we now briefly demonstrate its
applicability for a number of other quantum dynamic systems that have
been covered in \cite{Carlini 1,Morrison 1,Morrison 3}. These quantum
systems are of particular interest for quantum control, and can be
easily related to real physical systems that can be measured in the
laboratory. The relationship between the special unitary group SU(2)
and spin-orbit coupling, as in the Zeeman effect, is well known. We
can assume a constraint magnetic field $\tilde{F}=\lambda\hat{\sigma}_{z}$
with $\tilde{H}=\epsilon\hat{\sigma}_{+}+\epsilon^{\star}\hat{\sigma}_{-}$;
using the preceding analysis to solve the quantum brachistochrone
equation, we arrive at the unitary/optimal Hamiltonian pair below:
\begin{equation}
\left.\begin{array}{c}
\tilde{H}(t)=\left[\begin{array}{cc}
0 & e^{-it}\\
e^{+it} & 0
\end{array}\right]\\
\\
\hat{U}(t,s)=\left[\begin{array}{cc}
1 & 0\\
0 & e^{i(t-s)}
\end{array}\right]
\end{array}\right.
\end{equation}
where it is simple to show the isometric property $\tilde{H}(t)=\hat{U}(t,s)\tilde{H}(s)\hat{U}^{\dagger}(t,s)$
as required. We have normalised constant coefficients to unity for
the purposes of discussion. The preceding analytic technique improves
the speed of calculation, as the method of \cite{Carlini 1,Carlini 2}
requires a lengthy argument relating to the boundary conditions on
the state space for particular choices of input and target vectors.
The method presented completely avoids this issue and allows us to
find exact solutions on higher order spaces with strange geometric
features. For an example of a more exotic choice of dynamics, one
might consider a particle embedded in SU(3). This problem was analysed
in detail originally in \cite{Morrison 1} using techniques developed
in \cite{Carlini 1}. We present the unitary/Hamiltonian pair: 
\begin{equation}
\left.\begin{array}{c}
\tilde{H}(t)=\left[\begin{array}{ccc}
0 & \cos(t) & 0\\
\cos(t) & 0 & -ie^{-i\theta}\sin(t)\\
0 & +ie^{i\theta}\sin(t) & 0
\end{array}\right]\\
\\
\hat{U}(t,s)=\left[\begin{array}{ccc}
\cos(t-s) & 0 & -ie^{-i\theta}\sin(t-s)\\
0 & 1 & 0\\
ie^{i\theta}\sin(t-s) & 0 & \cos(t-s)
\end{array}\right]
\end{array}\right.
\end{equation}
Note that this expression for the unitary operator is extremely compact
compared to the derived expansions used originally in \cite{Morrison 1}
to calculate brachistochrones on SU(3). We are not required to evaluate
many difficult steps in order to arrive at the unitary transformations
of the state, once we have constructed suitable eigenvectors using
results established in \cite{Morrison 1} and solved the quantum brachistochrone
equation. Comparing these two matrices with the formula derived for
the relativistic electron, we immediately observe that the unitary
operator has a centre in the case of SU(3) problem against the results
from SU(2) and SU(4), as analysed in detail. As a matter of practical
use, the operators shown for SU(3) could quite easily be used in a
quantum computation/control scenario to implement some interesting
gates. For example, we can write down the transformation which takes
us into the eigenstate representation: 
\begin{equation}
\hat{Q}(t)=\left[\begin{array}{ccc}
\dfrac{1}{\sqrt{2}}\mathrm{\cos(}t) & -\dfrac{1}{\sqrt{2}}\cos(t) & ie^{-i\theta}\sin(t)\\
\dfrac{1}{\sqrt{2}} & \dfrac{1}{\sqrt{2}} & 0\\
\dfrac{i}{\sqrt{2}}e^{i\theta}\sin(t) & -\dfrac{i}{\sqrt{2}}e^{i\theta}\sin(t) & \cos(t)
\end{array}\right]
\end{equation}
and in particular
\begin{equation}
\hat{Q}(0)=\left[\begin{array}{ccc}
\dfrac{1}{\sqrt{2}} & -\dfrac{1}{\sqrt{2}} & 0\\
\dfrac{1}{\sqrt{2}} & \dfrac{1}{\sqrt{2}} & 0\\
0 & 0 & 1
\end{array}\right]
\end{equation}
which is in a suitable form for application within a quantum logic
schemata as a qutrit gate. Now, in contrast, the above two unitary
operators follow the equation $\hat{U}(t,s)=\hat{Q}(t)\hat{Q}^{\dagger}(s)$
against the difficulties experienced in the problem of the relativistic
electron, which has the property that $\hat{U}(t,s)=\hat{W}(t)\hat{W}^{-1}(s)$
. So, we can classify the types of physical systems by whether the
matrix of eigenvectors is unitary, and whether their time evolution
operator has a centre. There may be further emergent properties to
be discovered, especially for new groups and dynamical systems that
are amenable to this analysis.

\section{Discussion \& Future Directions}

We have shown in this paper how one might address the question of
relativistic electron dynamics from a perspective of time optimal
state control. The methods that we have outlined and applied to this
problem are indeed more general. One is free to move up, down or within
any quantum system and apply a similar formalism of constraints on
the system degrees of freedom that are accessible. The results should
be similar. Whether that is to be the case remains to be established
in general, apart from the examples we have demonstrated there are
remarkably few examples of these types of exactly solveable quantum
systems. Their mathematical nature is tightly tied to the symmetries
of the system we constrain the dynamics to evolve throughout; although
we do not specifically demand that the motion be periodic over the
time interval it has naturally emerged as a derivate property. This
is a distinct mathematical curiousity, with specific, testable consequences
for how we view quantum systems, and how we best go about planning
functional devices on a nanoscale.

It is a simple exercise to show Lorentz invariance of the formalism
developed for the SU(4) problem. We can simply transform the unitary
operator further under a Lorentz rotation in space and time. That
it is invariant is a consequence of the constraints of the problem.
We take particular note that the rest mass of the electron/positron
is a constant of motion. It is indeed fortunate, for if it were otherwise,
we would have significant scientific reason to invalidate the theory.
Some other interesting features of the dynamic system are immediately
apparent. In a deeper sense, this model of the spinning electron/positron
system gives validation to the concept of electronic indivisibility,
a predicted and measurable fact. We know, from experiment, that the
rest mass is the same for all electrons, and does not change. It seems
that, in this case, the spinning electron also holds no (extra) mass.
It is difficult to add mass to an electron. Further papers will address
the scattering theory of these types of systems using the anti-commuting
Hermitian variables we have found so useful in this paper. The extension
of this argument will be discussed; the observation of Lorentz invariance
shows that one may move to a situation of non-zero electromagnetic
potential using the gauge invariance principle. Whether this addresses
the question of constancy of charge for is to be demonstrated. One
would hope that this turns out to be the case.

The use of the Hamiltonian method to solve these types of questions,
while not unheard of, has enjoyed a sojourn in the field of quantum
electrodynamics. The primacy of the Lagrangian method, while enabling
calculation of certain types of scattering problems, has served to
obfuscate the true nature of the dynamics of electrons. Feynman \cite{Feynman 1}
writes that by \textquotedblleft ..forsaking the Hamiltonian method,
the wedding of relativity and quantum mechanics can be accomplished
most naturally\textquotedblright . It may be a matter of taste and
aesthetics, but one might respectfully disagree, given the ease of
calculation presented throughout this paper. The bringing together
of these seemingly disparate topics of scattering theory and time
optimal quantum control is sure to be a productive and fruitful enterprise.
The parallels to established results in quantum field theory are easily
seen and will be explored in future expositions.

We have addressed some outstanding questions in modern quantum mechanics
with this calculation. Feynman, in his paper on operator calculus
\cite{Feynman 2}, states that if ``..other operators are involved,
such as Pauli's spin operators or Dirac matrices which satisfy different
commutation rules, a complete reduction eliminating all the operators
is not nearly so easily affected....the amplitude for a single trajectory
is then a hypercomplex quantity in the algebra of the gamma or sigma
matrices\textquotedblright{} and then remarks further that \textquotedblleft ..not
much has been done with this expression... (It is suggestive that
the perhaps coordinates and the space-time they represent may in some
future theory be replaced completely by analysis of ordered quantities
in some hypercomplex algebra)''. Tomonaga \cite{Tomonaga} also comments
on the nature of time in his seminal paper on Lorentz invariant field
equations, stating that one \textquotedblleft ..sees that time plays
also here a role distinguished from x,y and z; also here a plane parallel
to the xyz-plane has a special significance. So we must in some way
remove this unsatisfactory feature of the theory\textquotedblright{}
and furthermore that the \textquotedblleft{} ..reason why the ordinary
formalism of the quantum field theory is so unsatisfactory lies in
the fact that one has built up this theory in the way which is too
much analogous to the ordinary non-relativistic mechanics.\textquotedblright{}
With this calculation, we have achieved a realisation of this hypercomplex
algebra, and it seems to serve particular utility within this application
of quantum state control. That it relates directly to the optimisation
of time, and the nature of time itself within quantum systems, is
an interesting answer to the questions both Feynman and Tomonaga raise
in their papers on quantum electrodynamics. 

Using the method we have presented, we are able to evaluate exact
solutions for time-ordered exponentials that have not been available
before, for explicitly time dependent quantum systems, while successfully
avoiding the difficulties of Dyson series, Magnus series, Lie-Trotter
expansions and the Baker-Campbell-Haussdorff approach. The problem
has been recast to one of diagonalisation into the set of initial
eigenstates, and multiplication by an exponential matrix. A clever
observation allows us to look at the unitary operator, akin to the
monodromy matrix in standard Floquet theory, as composed of one operator
moving forward in time and another moving backwards in time. This
transformation allows us to readily evaluate the system. All that
is relied upon is the form of the initial eigenvectors of the Hamiltonian
operator, and the way in which it evolves in time. For the non-trivial
systems examined so far, we have observed both periodic and constant
behaviour. Whether this turns out to be the case in general, i.e.
the Hamiltonian for these types of quantum systems is either periodic,
or constant, remains to be shown. There are an infinite amount of
differential equations that can be posed in this form, and we can't
claim to have solved them all. We have, however, established results
on all groups of matrices less than 4$\times$4. There are obviously
generalisations that can be made to spin chains and other systems
which might have a sensible continuous limit. The way forward seems
familiar, but is not well trodden.

From an experimental perspective, we can observe a certain consistency
in the time-dependent Hamiltonian systems we have developed in this
paper. We require, it appears, to be able to resonate with the device
using strong fields. As the interaction of the current with the device
is aperiodic within DC-type solid state devices, we propose the alternative
of making these resonant fields the driving force of the quantum state
rather than a source of error. Forms this might take include AC spin-tronics
on quantum dots, chemical control, NMR or other systems where time
dependent fields are easily introduced. Principal challenges to this
implementation may include the complex engineering task of rendering
a power source with stable driving oscillator current to exist within
the circuit design on a nano-scale. Differential analysers and computers
that run off alternating current exist on a macro scale, one learns
of such techniques in any standard class on AC circuit theory. Reduction
of AC components and retooling of the computational design and hardware
to deal with oscillatory signals on the nano-scale is a major engineering
task. Despite the challenges, the theory should be readily testable
in at least a scattering context and within spin-NMR type systems.

Various ways in which time optimal control theory may be extended
include the addition of non-linear constraints to the action principle.
When the Hamiltonian operator is varied, these will result in additional
terms in the quantum brachistochrone equation, and may describe some
interesting non-linear dynamics. It may or may not be consistent with
current quantum mechanics, as thus far we have only required linear
constraints in order to describe the relevant physics, however, it
is likely to be of interest to those in the dynamical systems field.
Much remains to be done regarding the implementation of numeric methods
to address these types of coupled non-linear ordinary differential
equations. The problems will not scale well as dimension increases,
so a good understanding of the nature of the underlying groups that
drive the dynamics is likely to be a place of fruitful endeavour. 

Finally, we state that the extension of this type of time optimal
quantum control problem to the continuum is one of pressing interest.
The equations, being cast in the form of finite matrices, are only
true for finite dimensional objects. One would hope that there would
be a continuous version of the quantum brachistochrone which would
be the result of some limit of matrices extending to infinity. This
hope, while likely misplaced, should be explored further. To date
this has not been carried out. The known relationships between continuous
groups and infinite dimensional matrices may be of some use here,
however the answer to this question remains outstanding. As we have
gone up in dimension from SU(2), through SU(3) and finally to the
SU(4) examples examined in this paper, we have observed the emergence
of several completely different types of unitary systems. Whether
a sensible limit at the point of infinity exists is an open question.

\bibliographystyle{alpha} \bibliographystyle{alpha}
\bibliography{abbrv,textbooks,QuantumComputation,stallisEntropy}

\begin{thebibliography}{References}
\bibitem[1]{Almeida}Almeida, E. d., \& Sharma, S. S. (2004). \textit{SU(3)
model description of Be isotopes}. Brazilian journal of physics, 34(3A),
962-965.

\bibitem[2]{Asarin}Asarin, E., \& Maler, O. (1999). \textit{As soon
as possible: Time optimal control for timed automata}. Paper presented
at the International Workshop on Hybrid Systems: Computation and Control.

\bibitem[3]{Bobrow}Bobrow, J. E., Dubowsky, S., \& Gibson, J. S.
(1985). \textit{Time-Optimal Control of Robotic Manipulators Along
Specified Paths}\emph{.} The International Journal of Robotics Research,
4(3), 3\textendash 17.

\bibitem[4]{Brockett 1}Brockett, R. W. (1982). \textit{Control theory
and singular Riemannian geometry}\emph{.} New directions in applied
mathematics (pp. 11-27): Springer. 

\bibitem[5]{Brockett 2}Brockett, R. W. (2007). \textit{Optimal control
of the Liouville equation}. AMS IP Studies in Advanced Mathematics,
39, 23. 

\bibitem[6]{Carlini 1}Carlini, A., Hosoya, A., Koike, T., \& Okudaira,
Y. (2006). \textit{Time-optimal quantum evolution}. Physical review
letters, 96(6), 060503.

\bibitem[7]{Carlini 2}Carlini, A., Hosoya, A., Koike, T., \& Okudaira,
Y. (2007). \textit{Time-optimal unitary operations}. Physical Review
A, 75(4), 042308. 

\bibitem[8]{Carlini 3}Carlini, A., Hosoya, A., Koike, T., \& Okudaira,
Y. (2008). \textit{Time optimal quantum evolution of mixed states}.
Journal of Physics A: Mathematical and Theoretical, 41(4), 045303. 

\bibitem[9]{Carlini 4}Carlini, A., Hosoya, A., Koike, T., \& Okudaira,
Y. (2011). \textit{Time-optimal CNOT between indirectly coupled qubits
in a linear Ising chain}. Journal of Physics A: Mathematical and Theoretical,
44(14), 145302. 

\bibitem[10]{Carlini 5}Carlini, A., \& Koike, T. (2012). \textit{Time-optimal
transfer of coherence}. Physical Review A, 86(5), 054302. 

\bibitem[11]{Chen}Chen, Q.-M., Wu, R.-B., Zhang, T.-M., \& Rabitz,
H. (2015). \textit{Near-time-optimal control for quantum systems}.
Physical Review A, 92(6), 063415.

\bibitem[12]{Dirac}Dirac, P. A. M. (1936). \textit{Relativistic wave
equations}. Proceedings of the Royal Society of London. Series A-Mathematical
and Physical Sciences, 155(886), 447-459.

\bibitem[13]{Fattorini}Fattorini, H. (1964). \textit{Time-optimal
control of solutions of operational differenital equations}. Journal
of the Society for Industrial and Applied Mathematics, Series A: Control,
2(1), 54-59. 

\bibitem[14]{Feynman 1}Feynman, R. P. (1949). \textit{Space-time
approach to quantum electrodynamics}. Physical Review, 76(6), 769. 

\bibitem[15]{Feynman 2}Feynman, R. P. (1951). \textit{An operator
calculus having applications in quantum electrodynamics}. Physical
Review, 84(1), 108. 

\bibitem[16]{Floquet}Floquet, G. (1883). \textit{Sur les équations
différentielles linéaires à coefficients périodiques}. Annales Scientifiques
de l'École Normale Supérieure, 12, 47-88.

\bibitem[17]{Friesz}Friesz, T., Luque, J., Tobin, R., Wie B., (1989).
\textit{Dynamic Network Traffic Assignment Considered as a Continuous
Time Optimal Control Problem}. Operations Research 37(6):893-901

\bibitem[18]{Gross}Gross, P., Neuhauser, D., \& Rabitz, H. (1992).
\textit{Optimal control of curve crossing systems}. The Journal of
chemical physics, 96(4), 2834-2845. 

\bibitem[19]{Jirari}Jirari, H., Hekking, F. W., \& Buisson, O. (2009).
\textit{Optimal control of superconducting N-level quantum systems}.
EPL (Europhysics Letters), 87(2), 28004. 

\bibitem[20]{Kehlet}Kehlet, C. T., Sivertsen, A. C., Bjerring, M.,
Reiss, T. O., Khaneja, N., Glaser, S. J., \& Nielsen, N. C. (2004).
\textit{Improving solid-state NMR dipolar recoupling by optimal control}.
Journal of the American Chemical Society, 126(33), 10202-10203. 

\bibitem[21]{Khaneja 1}Khaneja, N., Brockett, R., \& Glaser, S. J.
(2001). \textit{Time optimal control in spin systems}. Physical Review
A, 63(3), 032308. 

\bibitem[22]{Khaneja 2}Khaneja, N., Glaser, S. J., \& Brockett, R.
(2002). \textit{Sub-Riemannian geometry and time optimal control of
three spin systems: quantum gates and coherence transfer}. Physical
Review A, 65(3), 032301. 

\bibitem[23]{Khaneja 3}Khaneja, N., Reiss, T., Kehlet, C., Schulte-Herbrüggen,
T., \& Glaser, S. J. (2005). \textit{Optimal control of coupled spin
dynamics: design of NMR pulse sequences by gradient ascent algorithms}.
Journal of magnetic resonance, 172(2), 296-305. 

\bibitem[24]{Lapert}Lapert, M., Zhang, Y., Braun, M., Glaser, S.,
\& Sugny, D. (2010). \textit{Singular extremals for the time-optimal
control of dissipative spin 1 2 particles}. Physical review letters,
104(8), 083001.

\bibitem[25]{Leclerc}Leclerc, M. (2006).\emph{ }\textit{Hermitian
Dirac Hamiltonian in the time-dependent gravitational field}. Classical
and Quantum Gravity, 23(12), 4013. 

\bibitem[26]{Lee}Lee, H., Teo, K., Rehbock, V., \& Jennings, L. (1997).
\textit{Control parametrization enhancing technique for time optimal
control problems}. Dynamic Systems and Applications, 6, 243-262. 

\bibitem[27]{Li}Li, J., Yang, X., Peng, X., \& Sun, C.-P. (2017).
\textit{Hybrid quantum-classical approach to quantum optimal control}.
Physical review letters, 118(15), 150503. 

\bibitem[28]{Mandal}Mandal, B. P., \& Gupta, S. (2010). \textit{Pseudo-hermitian
interactions in Dirac theory: Examples}. Modern Physics Letters A,
25(20), 1723-1732. 

\bibitem[29]{Mayne}Mayne, D. Q., \& Schroeder, W. (1997). \textit{Robust
time-optimal control of constrained linear systems}. Automatica, 33(12),
2103-2118. 

\bibitem[30]{Morrison 1}Morrison, P. G. (2008), \textit{Time optimal
quantum state control}. Thesis, Centre for Quantum Computer Technology,
Macquarie University.

\bibitem[31]{Morrison 2}Morrison, P. G. (2012), \textit{Time dependent
biqubits}. Australian Conference on Optical Fibre Technology/AIP Proceedings,
2012: 1-4. 

\bibitem[32]{Morrison 3}Morrison, P. G. (2012), \textit{Time dependent
quantum mechanics}. arxiv.org/abs/1210.6977

\bibitem[33]{Mukherjee}Mukherjee, V., Carlini, A., Mari, A., Caneva,
T., Montangero, S., Calarco, T., Giovannetti, V. (2013). \textit{Speeding
up and slowing down the relaxation of a qubit by optimal control}.
Physical Review A, 88(6), 062326. 

\bibitem[34]{Palao} Palao, J. P., \& Kosloff, R. (2002). \textit{Quantum
computing by an optimal control algorithm for unitary transformations}.
Physical review letters, 89(18), 188301. 

\bibitem[35]{Redkov}Red'kov, V. M., Bogush, A. A., \& Tokarevskaya,
N. G. (2008). \textit{On Parametrization of the Linear GL(4,C) and
Unitary SU(4) Groups in Terms of Dirac Matrices}. Symmetry, Integrability
and Geometry: Methods and Applications, 4(0), 21-46. 

\bibitem[36]{Rodionov}Rodionov, V. N. (2015). \textit{Exact solutions
for non-Hermitian Dirac-Pauli equation in an intensive magnetic field}.
Physica Scripta, 90(4), 045302. 

\bibitem[37]{Santoro}Santoro, G., Introduction to Floquet- Lecture
notes, SISSA Trieste

\bibitem[38]{Schulte-Herbruggen}Schulte-Herbrüggen, T., Spörl, A.,
Khaneja, N., \& Glaser, S. (2005). \textit{Optimal control-based efficient
synthesis of building blocks of quantum algorithms: A perspective
from network complexity towards time complexity}. Physical Review
A, 72(4), 042331. 

\bibitem[39]{Shi}Shi, Z.-Q. (2011). \textit{Analytical Study of the
Spin Projection Operator}\emph{.} arXiv preprint arXiv:1101.0481. 

\bibitem[40]{Sklarz}Sklarz, S. E., Tannor, D. J., \& Khaneja, N.
(2004). \textit{Optimal control of quantum dissipative dynamics: Analytic
solution for cooling the three-level lambda system}. Physical Review
A, 69(5), 053408. 

\bibitem[41]{Somloi}Somlói, J., Kazakov, V. A., \& Tannor, D. J.
(1993). \textit{Controlled dissociation of I2 via optical transitions
between the X and B electronic states}. Chemical physics, 172(1),
85-98. 

\bibitem[42]{Tomonaga}Tomonaga, S.-i. (1946).\emph{ }\textit{On a
relativistically invariant formulation of the quantum theory of wave
fields}. Progress of Theoretical Physics, 1(2), 27-42. 

\bibitem[43]{Vosegaard}Vosegaard, T., Kehlet, C., Khaneja, N., Glaser,
S. J., \& Nielsen, N. C. (2005). \textit{Improved excitation schemes
for multiple-quantum magic-angle spinning for quadrupolar nuclei designed
using optimal control theory}. Journal of the American Chemical Society,
127(40), 13768-13769. 

\bibitem[44]{Wang}Wang, X., Allegra, M., Jacobs, K., Lloyd, S., Lupo,
C., \& Mohseni, M. (2015). \textit{Quantum brachistochrone curves
as geodesics: Obtaining accurate minimum-time protocols for the control
of quantum systems}. Physical review letters, 114(17), 170501. \end{thebibliography}

\end{document}